# Scientometrics: Untangling the topics

Ádám Szántó-Várnagy [1], Péter Pollner [2,3], Tamás Vicsek [1,2,3], Illés J. Farkas [2,3,*]

*Measuring science is based on comparing articles to similar others. However, keyword-based groups of thematically similar articles are dominantly small. These small sizes keep the statistical errors of comparisons high. With the growing availability of bibliographic data such statistical errors can be reduced by merging methods of thematic grouping, citation networks and keyword co-usage.*

Pieces of our collective human scientific knowledge are constantly defined and modified through our global scientific communication. The most common units of this process are publications, also called articles or papers. These units (i) provide "road signs" for newcomers to a field and (ii) allow the scientific community to steer its work toward consensus-based goals given the available resources. *Due to the size of science automated measurements are necessary* to achieve these two goals. In particular, the steering aspect involves decisions about manuscript acceptance and science funding, which includes even jobs of scientists. Thus, it seems reasonable to move to the public domain not only scientometric algorithms but also bibliographic data[1]. With more data in the public domain our current assumptions about the data itself may be challenged.

To measure science, one needs to measure the scientific communication process, which is *a network of articles (nodes) connected by citations (directed links) and tagged with article keywords*. Most current scientific metrics are built on article-level metrics[2-13] (ALMs) and the most common ALM is the (total) citation number. The citation number – similarly to other mention-counting ALMs – has the following major properties. First, there are more publications every year (Fig.1a) and the number of references per publication is growing too (Fig.1b). Second, papers with an earlier publication date have had until now more time to receive citations. Third, the citation count by itself blanks out citation context[2], which includes citing paper quality. In summary, the citation number tends to favor papers that appeared close (in time and topic) to the origins of large and still active research areas. Improvements to the citation number focus on (i) the topic and (ii) quality of citing papers, (iii) the time of publication and (iv) the current state of a paper's research area.

The research areas (topics) of a paper are shown by its keywords. Even though most papers have more than one keyword (Fig.1c), within a small group of papers total citation numbers can be manually adjusted. Scaling up this manual comparison leads to *the automated classification of all papers into research areas*[3,4] and to the normalization of any paper's citation number based on the total number of papers and citations in its field(s) and publication year[5,6,7]. To include citing paper quality, the PageRank algorithm[8] identifies publications with highly cited "descendants". To filter out inactive fields of research the CiteRank[9] and Discounted Cumulated Impact[10] (DCI) indexes include the ageing of scientific content, while FutureRank[11] and the Minimal Citation (MiC) model[12] identify "rising star" publications by estimating future citation numbers. These and other quantitative tools are necessary for both learning and science-related decisions.

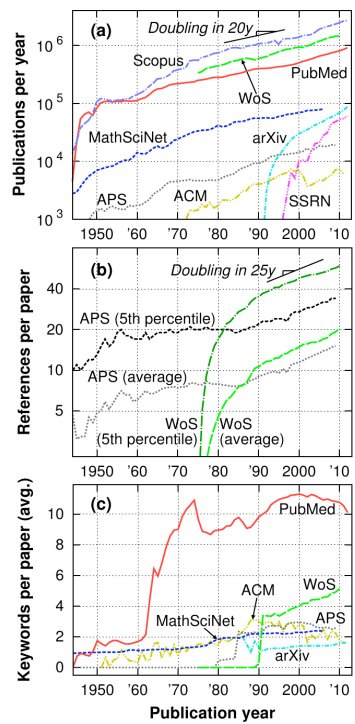

**Figure 1** (color). Scientific publication statistics by year from the ACM Digital Library (ACM), the American Physical Society (APS), the arXiv, MathSciNet, PubMed, Scopus, the Social Science Research Network (SSRN) and the Web of Science (WoS). Scopus data assign January 1 to previous year. WoS data licensed by EU ERC COLLMOT.

Both major applications of measuring science (i.e., learning and decisions) compare papers, individuals, groups or institutions to similar others. Note that these comparisons are all built on comparing papers (articles). A comparison of articles assumes that we can assign each to one or more article sets that are characterized by averages (medians) taken over the given set. In fact, the existence of such homogeneous article groups is *an unspoken axiom in scientometrics:* it is widely assumed that all scientific articles can be assigned to thematically homogeneous groups of articles. To keep statistical errors low these groups need to be large.

With keywords the least and most stringent conditions of thematic similarity in a group of papers are that (a) all papers share at least one keyword and (b) all papers have the exact same keyword list. Figures 2a and 2b show that the distribution of the sizes of such article groups decreases (at medium and large group sizes) faster than a power-law with slope –1. With simple math this means that the probability for a paper to belong to a group drops with the group's size faster than a power-law with exponent 0, which is a constant. So a paper is more likely to belong to a small group than to a large group. Moreover, if only papers with similar publication dates are allowed in a thematic group, then group sizes are further reduced. In summary, the above unspoken axiom implies that instead of homogeneous large groups of papers *science is dominated by homogeneous small groups of papers*.

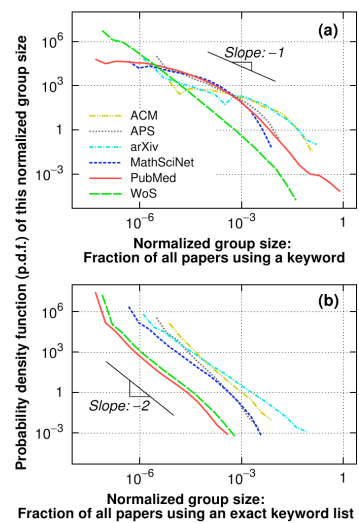

**Figure 2** (color). Publication databases cover different scientific fields with different methods. Nonetheless, they show similar distributions for the fraction of all papers using (a) a keyword or (b) an exact keyword list. Logarithmic binning is applied.

Two consequences of the dominance of small article groups are that (i) a

keyword-based comparison of articles with thematically similar others keeps *statistical errors high* (with all analyzed keyword schemes) and (ii) *these errors propagate from article-level metrics to all other metrics*. The growing availability of bibliographic data may reduce this type of statistical error. It allows now the integration of content-based keyword assignment schemes with citation networks [13] and the network of keywords as defined by their joint usage on publications (Fig.2b). We point out that in Figure 2 keywords provided by authors (*e.g.*, APS PACS terms or arXiv categories) and keywords assigned by databases (*e.g.*, PubMed MeSH terms or WoS KeyWordPlus terms) show similar distributions. This and other universal properties[5] of large-scale bibliographic data may provide *more precise standards* for quantifying scientific contributions.

Ádám Szántó-Várnagy, Péter Pollner, Tamás Vicsek, Illés J. Farkas
[1] Institute of Physics, Eotvos University, Budapest, Hungary
[2] MTA-ELTE Statistical and Biological Physics Group, Budapest, Hungary
[3] Regional Knowledge Centre, ELTE Faculty of Sciences, by EU-ESF Támop 4.2.2.C-11/1/KONV-2012-0013, Székesfehérvár, Hungary.
**\* Corresponding author**
E-mail: fij@elte.hu